# Analysis of 24-Hour Ambulatory Blood Pressure Monitoring Data using Orthonormal Polynomials in the Linear Mixed Model

**Lloyd J. Edwards[1*] and Sean L. Simpson[2]**

[1]Department of Biostatistics, University of North Carolina at Chapel Hill,

Chapel Hill, North Carolina 27599-7420

[2]Department of Biostatistical Sciences, Wake Forest School of Medicine,

Winston-Salem, NC 27157-1063

[*]*email:* edwards@bios.unc.edu

SUMMARY. The use of 24-hour ambulatory blood pressure monitoring (ABPM) in clinical practice and observational epidemiological studies has grown considerably in the past 25 years. ABPM is a very effective technique for assessing biological, environmental, and drug effects on blood pressure. In order to enhance the effectiveness of ABPM for clinical and observational research studies via analytical and graphical results, developing alternative data analysis approaches are important. The linear mixed model for the analysis of longitudinal data is particularly well-suited for the estimation of, inference about, and interpretation of both population and subject-specific trajectories for ABPM data. Subject-specific trajectories are of great importance in ABPM studies, especially in clinical research, but little emphasis has been placed on this dimension of the problem in the statistical analyses of the data. We propose using a linear mixed model with orthonormal polynomials across time in both the fixed and random effects to analyze ABPM data. Orthonormal polynomials in the linear mixed model may be used to develop model-based, subject-specific 24-hour ABPM correlates of cardiovascular disease outcomes. We demonstrate the proposed analysis technique using data from the Dietary Approaches to Stop Hypertension (DASH) study, a multicenter, randomized, parallel arm feeding study that tested the effects of dietary patterns on blood pressure.

KEY WORDS: Cubic Spline, DASH Study, Graphical Display, Hypertension, Longitudinal, Model Selection, Orthogonal Polynomials

## 1. Introduction - Motivation and Background

Ambulatory blood pressure monitoring (ABPM) is a powerful research tool for examining blood pressure (BP) variability and the physiologic and environmental factors that affect BP.[1] The evidence that ABPM gives information over and above conventional blood pressure measurement (CBPM) has been growing steadily over the past 25 years and the technique is now accepted as being indispensable to good clinical practice.[2-3] ABPM is a non-invasive technique in which a standard cuff, attached to a lightweight, portable data recording unit, is placed around the upper arm and inflated at regular preset intervals during a 24 hour time period.

Traditionally, clinicians have used office blood pressure measurements as the preferred method of monitoring blood pressure and consequently the method of diagnosing hypertension. Unfortunately, physician's office blood pressure measurements often can give rise to what is termed "white-coat hypertension", that is, an artificially high blood pressure reading.[4] ABPM provides a profile of blood pressure away from the medical environment, thereby allowing identification of individuals with a white coat response.[5] It also provides several other important advantages discussed by O'Brien[5]: ABPM can demonstrate the efficacy of antihypertensive medication over a 24 hour period rather than making a decision based on one or a few CBPMs confined to a short period of the diurnal cycle; ABPM can identify patients whose blood pressure does not reduce at night-time (the non-dippers) who are probably at high risk for a number of conditions; and the technique can demonstrate a number of patterns of blood pressure behavior that may be relevant to clinical management (isolated systolic hypertension, hypotension, dipping and non-dipping, etc.).

As noted by O'Brien[5], recent longitudinal studies have shown that ABPM is a much stronger predictor of cardiovascular morbidity and mortality than CBPM. The importance of the technique is further evidenced by the fact that the Centers for Medicare and Medicaid Services (CMS) in the US have approved ABPM for reimbursement. O'Brien[2,5] concluded that ABPM should be available to all hypertensive patients given its international acceptance as an indispensable tool for patients with established and suspected hypertension. This recommendation has important implications for clinical practice. Among the questions that O'Brien[5] posed were: 1. how should the data be presented?; 2. how best can doctors and nurses unfamiliar with the technique be educated in its use and interpretation of the data?

Lambert et al.[6] noted that there has been relatively little work done on the longitudinal analysis of 24-hour ABPM data. Though some authors have attempted to address the issue, there is not a generally accepted 'standard' method of analyzing 24-hour ABPM. Some analyses of ABPM data have used means and/or medians, either obtaining means over the entire 24-hour period or obtaining means over defined intervals of the 24-hour period, for example daytime and nighttime intervals.[7-9] Other analyses have stressed the use of traditional simple or extended cosinor models.[10-12] Some authors have proposed smoothing splines or other smoothing techniques for the analysis of 24-hour ABPM data .[13-15] Jaccard and Wan[16] proposed using cross-sectional pooled time series designs for the analysis of 24-hour ABPM data. Schwartz et al.[17] proposed a very limited variance-components model which is equivalent to a mixed model using very simple explanatory factors (time is not explicitly used as a factor). Selwyn and DiFranco[18] demonstrated the use of the general linear mixed model as an approach to analyzing 24-hour ABPM data. Typically the emphasis of 24-hour ABPM data analyses are on between-subject effects (fixed effects) and less emphasis is placed on within-subject effects (random effects).[6]

We propose using a linear mixed model with orthonormal polynomials in both the fixed and random effects as an effective approach to analyzing 24-hour ABPM data, particularly when subject-specific trajectories are of primary emphasis. The general linear mixed model[19-20] has become a very powerful statistical tool in the analysis of longitudinal data with continuous outcomes in both clinical and non-clinical studies.[21-23] The linear mixed model for the analysis of longitudinal data is particularly well-suited for the estimation of, inference about, and interpretation of both population and subject-specific trajectories for ABPM data. Subject-specific trajectories are of great importance in ABPM studies but little emphasis has been placed on this dimension of the problem in data analyses. Using orthonormal polynomials provides the ability to address the nonlinear trajectories represented by individual subjects' trajectories with the same complexity in which the mean model (fixed effects) are addressed. Graphical presentations are proposed that help to address questions posed by O'Brien[5] regarding how the data should be presented. We demonstrate the proposed technique using data from the Dietary Approaches to Stop Hypertension (DASH) study, a multicenter, randomized, parallel arm feeding study that tested the effects of dietary patterns on blood pressure.

In Section 2, a description of the DASH trial is given. In Section 3, the basic linear mixed model formulation and the details for the orthonormal polynomial statistical methods are presented. We discuss the results of applying the models to the DASH data in Section 4. Section 5 discusses the new graphical displays created, and a summary and conclusions are given in Section 6.

## 2. Description of Dietary Approaches to Stop Hypertension (DASH) Clinical Trial

The Dietary Approaches to Stop Hypertension (DASH) trial was a multicenter, randomized, parallel arm feeding study that tested the effects of dietary patterns on blood pressure (Appel et al 1997 - 9). The three diets were a control diet (low in fruits, vegetables, and dairy products, with a fat content typical of the average diet in the United States), a diet rich in fruits and vegetables (a diet similar to the control except it provided more fruits and vegetables and fewer snacks and sweets), and a combination diet rich in fruits, vegetables, and low-fat dairy foods and reduced in saturated fat, total fat, and cholesterol. The combination diet will be subsequently referred to as the DASH diet.

Participants were healthy, community-dwelling adults 22 years of age or older who were not taking antihypertensive medication. Each subject had an average systolic blood pressure of less than 160 mm Hg and a diastolic blood pressure of 80 to 95 mm Hg (mean of six measurements across three screening visits). Study subjects were enrolled sequentially in groups; the first group began the run-in phase of the trial in September 1994, and the fifth and last group started in January 1996.

For each group, data were collected during three phases (screening, run-in, and intervention). Run-in was a three-week period in which all participants were fed the control diet. Toward the end of run-in, 24-hour ambulatory blood pressure monitoring was obtained once. This constituted the "baseline" ABPM reading. During the third week, participants were randomized to one of three diets. Intervention was an eight-week period in which participants were fed their assigned diets. During the last two weeks, one 24-hour ambulatory blood pressure monitoring was obtained (end-of-intervention ABPM).

ABPM was attempted on the 362 participants enrolled in groups 2-5. We use 357 subjects for baseline 24-hour ABPM data analysis, including incomplete data. We also use the 288 subjects who had complete data at baseline and the 287 subjects who had complete data during the intervention for our AUC example.

## 3. The Linear Mixed Model and Orthonormal Polynomials

As discussed in Section 1.2, several statistical methods have been proposed for the analysis of 24-hour ABPM data. In this paper, we consider the linear mixed model with orthonormal polynomials with degree of polynomial ranging from 4 to 9.

With $N$ independent sampling units (often *persons* in practice), the linear mixed model for person $i$ may be written

$$\boldsymbol{y}_i = \boldsymbol{X}_i\boldsymbol{\beta} + \boldsymbol{Z}_i\boldsymbol{d}_i + \boldsymbol{e}_i. \tag{1}$$

Here, $\boldsymbol{y}_i$ is a $p_i \times 1$ vector of observations on person $i$; $\boldsymbol{X}_i$ is a $p_i \times q$ known, constant design matrix for person $i$, with full column rank $q$ while $\boldsymbol{\beta}$ is a $q \times 1$ vector of unknown, constant, population parameters. Also $\boldsymbol{Z}_i$ is a $p_i \times m$ known, constant design matrix with rank $m$ for person $i$ corresponding to the $m \times 1$ vector of unknown random effects $\boldsymbol{d}_i$, while $\boldsymbol{e}_i$ is a $p_i \times 1$ vector of unknown random errors. Gaussian $\boldsymbol{d}_i$ and $\boldsymbol{e}_i$ are independent with mean $\boldsymbol{0}$ and

$$\mathcal{V}\left(\begin{bmatrix} d_i \\ e_i \end{bmatrix}\right) = \begin{bmatrix} \Sigma_{di}(\tau_d) & 0 \\ 0 & \Sigma_{ei}(\tau_e) \end{bmatrix}. \tag{2}$$

Here $\mathcal{V}(\,\cdot\,)$ is the covariance operator, while both $\Sigma_{di}(\tau_d)$ and $\Sigma_{ei}(\tau_e)$ are positive-definite, symmetric covariance matrices. Therefore $\mathcal{V}(y_i)$ may be written $\Sigma_i = Z_i \Sigma_{di}(\tau_d) Z_i' + \Sigma_{ei}(\tau_e)$. We assume that $\Sigma_i$ can be characterized by a finite set of parameters represented by an $r \times 1$ vector $\tau$ which consists of the unique parameters in $\tau_d$ and $\tau_e$. Throughout $n = \sum_{i=1}^{N} p_i$.

We may also need to refer to a stacked data formulation of model (0) given by

$$y_s = X_s \beta + Z_s d_s + e_s, \tag{3}$$

with $y_s = [\,y_1' \;\cdots\; y_N'\,]'$, $X_s = [\,X_1' \;\cdots\; X_N'\,]'$, $Z_s = \mathrm{diag}(Z_1, \cdots, Z_N)$, $d_s = [\,d_1' \;\cdots\; d_N'\,]'$, and $e_s = [\,e_1' \;\cdots\; e_N'\,]'$. Here $d_s \sim \mathcal{N}_{Nm}[0, \Sigma_{di}(\tau_d) \otimes I_N]$ and $e_s \sim \mathcal{N}_n(0, \Sigma_{es})$ for $\Sigma_{es} = \mathrm{diag}[\Sigma_{e1}(\tau_e), \cdots, \Sigma_{eN}(\tau_e)]$. In turn $y_s \sim \mathcal{N}_n(X_s \beta, \Sigma_s)$ with $\Sigma_s = \mathcal{V}(y_s) = \mathrm{diag}(\Sigma_1, \cdots, \Sigma_N)$.

The advantage of reducing bias in covariance estimation has made restricted maximum likelihood (REML) estimation very popular for the linear mixed model. Given our focus on variance estimates, all parameter estimates in this paper are done using REML. However, the formulations also apply to computations based on maximum likelihood estimates.

*3.1 Why Orthonormal Polynomials?*

Polynomial models play a prominent role in mathematics. In mathematical analysis, the Weierstrass approximation theorem[24] states that every continuous function defined on a closed interval [a,b] can be uniformly approximated as closely as desired by a polynomial function. Because polynomials are among the simplest functions, and because computers can directly evaluate polynomials, this theorem has both practical and theoretical relevance, especially in polynomial interpolation. The Extendend Mean Value Theorem[24] is a remarkable theorem usually associated with Brook Taylor (1685-1731) that enables us to approximate various complicated functions by much simpler natural polynomial functions. Although a power series expansion cannot *always* be obtained, most of the familiar functions in calculus can be represented as a sum of a convergent power series. For any individual, blood pressure as a function of time on the interval [1,24] can be assumed to be a continuous function that meets the regularity conditions posed by the Weierstrass approximation theorem and the Extendend Mean Value Theorem.

Polynomials are commonly used to describe curved relationships in statistical models where a model must be developed empirically. Due to the heterogeniety of 24-hour ABPM trajectories among subjects, polynomials can be very effective statistical tools for empirically developing models for subject-specific trajectories. The goal of polynomial regression is to model a non-linear relationship between the independent and dependent variables (technically, between the independent variable and the conditional mean of the dependent variable). An advantage of traditional polynomial regression is that the inferential framework of the linear mixed model can be used.

The use of natural polynomials for the analysis of 24-hour ABPM recordings poses problems, particularly when modeling subject-specific trajectories. The problems include the existence of very large values in the fixed and random effects design matrices, for example $24^3 = 13,824$ and $24^9 = 2.64$ x $10^{12}$; problems of multicollinearity since $x$, $x^2$,…., $x^q$ are correlated; and mixed model software convergence problems when used in random effects. The lack of convergence in mixed model software can be especially acute when employing natural polynomials. In most cases it is caused by the multicollinearity and large values present in the random effects and their resulting effect on the estimation of the random effects covariance $\Sigma_{di}(\tau_d)$. The typical solution is to reduce the number of random effects, but this results in an increasing lack of fit for the subject-specific trajectories. The use of orthonormal polynomials helps alleviate this problem and hence allows for more accurate modeling of the heterogeneity in subject-specific 24-hour ABPM trajectories than natural polynomials.

Orthonormal polynomial predictors in linear regression are transformations of the natural polynomials that provide a new set of predictors that meet the following criteria: 1. The new predictors contain the same information as the original set; 2. The new predictors are linear combinations of the original natural polynomials; 3. The new columns of predictors all have mean zero (except for the constant term); 4. The new columns of predictors are all mutually orthogonal. Combining (3) and (4) implies that the new predictors (except the constant) are mutually uncorrelated. The first new column captures the information in the intercept, the second captures all of the linear term information adjusted for the intercept, the third captures quadratic information above and beyond the linear and intercept information and so forth.

In the case of 24-hour ABPM trajectories, the main interest is in the shape of the trajectories that the orthonormal polynomials produce collectively and not in any one single polynomial component. Two big advantages of using orthonormal polynomials are that a) the problem of multicollinearity is greatly reduced; and b) the orthonormal polynomials are "normalized" and bounded in the interval $(-1, 1)$. Thus, orthonormal polynomials have few problems with rounding error or very large (small) regression coefficients.

Using the linear mixed model with orthonormal polynomials, subject-specific 24-hour ABPM trajectories can be represented and parameterized, within- and between- treatment comparisons based on individual trajectories can be made, and discrimination between different trajectories can be obtained. The linear mixed model can account for circadian rhythms, subject effects, and treatment effects.

*3.2 The Case of Complete, Balanced Data with Equally Spaced Observations*

For the analysis of 24-hour ABPM data, we advocate the use of a linear mixed model with orthonormal polynomials used in both the fixed and random effects, particularly when an emphasis is placed on accurate modeling of subject-specific parameters. Here, we assume that the degree of the polynomials used in the fixed effects is equal to that used in the random effects. For demonstration purposes, we first start with a linear mixed model with only time effects and intercept, but no additional covariates. We assume that time is equally spaced 1-hour intervals ranging from 1 to 24 and that each subject has all 24 observations (complete and balanced). We have

$$\boldsymbol{y}_i = \boldsymbol{X}_i^*\boldsymbol{\beta} + \boldsymbol{Z}_i^*\boldsymbol{d}_i + \boldsymbol{e}_i \,. \tag{4}$$

Here, $\boldsymbol{y}_i$ is as before, a $p_i \times 1$ vector of observations on person $i$; $\boldsymbol{X}_i^*$ is a $p_i \times q$ fixed effects design matrix of orthonormal polynomials for person $i$; $\boldsymbol{Z}_i^*$ is a $p_i \times q$ random design matrix of orthonormal polynomials with rank $q$ for person $i$. Gaussian $\boldsymbol{d}_i$ and $\boldsymbol{e}_i$ are independent with mean $\boldsymbol{0}$ and variance given in equation 2.

Because the polynomials are orthogonal, mathematically we can model $\boldsymbol{\Sigma}_{di}(\boldsymbol{\tau}_d)$ as diagonal with heterogeneous diagonal elements, i.e., $\boldsymbol{\Sigma}_{di}(\boldsymbol{\tau}_d) = \text{diag}(\boldsymbol{\tau}_d)$, where $\boldsymbol{\tau}_d = (\sigma_0^2, \sigma_1^2, \ldots, \sigma_{q-1}^2)'$ is the $q \times 1$ vector of variances for each element of the random effects vector $\boldsymbol{d}_i$. Under this assumption, $\boldsymbol{\Sigma}_i = \boldsymbol{Z}_i^* \text{diag}(\boldsymbol{\tau}_d) \boldsymbol{Z}_i^{*\prime} + \sigma_e^2 \boldsymbol{I}_{p_i}$. This greatly simplifies the complexity of the random effects covariance and thus the modeling of $\boldsymbol{\mathcal{V}}(\boldsymbol{y}_i)$. The number of parameters in $\boldsymbol{\Sigma}_{di}(\boldsymbol{\tau}_d)$ is $q$, the number of diagonal elements. However, in actual data analysis, the best fitting $\widehat{\boldsymbol{\Sigma}}_{di}(\widehat{\boldsymbol{\tau}}_d)$ may be achieved by using an unstructured $\boldsymbol{\Sigma}_{di}(\boldsymbol{\tau}_d)$, where the collinearity of the elements of $\boldsymbol{\Sigma}_{di}(\boldsymbol{\tau}_d)$ is greatly reduced in comparison to a natural polynomials fit. These empirical covariances are present in the orthonormal fits due to the lack of fit of the fixed effects, $\boldsymbol{X}_i^* \widehat{\boldsymbol{\beta}}$, in equation 2. We address this issue in the example data.

The maximum degree orthonormal polynomial that we consider for the DASH data is a 9th degree. Our goal of modeling the nonlinear ABPM trajectories without overfitting the data along with our prior work and experience analyzing DASH data guided us to begin with a maximum model with a polynomial degree between 6 and 9.[25-26] We opted for the latter given that the 9th degree polynomial was significant in the SBP model (though not in the DBP model). Simpson and Edwards[26] used a fixed-effects only model for the DASH data and the order of the orthonormal polynomial was chosen to be 6 when considering additional covariates and aiming for parsimony. However, for the example analysis in this paper, we also consider a range of polynomials between 4 and 9 with a focus on the 9th degree model since it is the largest polynomial considered. We do not consider additional covariates but we discuss how the covariates can be included in the model.

In the 9th degree model, we estimate $q = 10$ fixed effects (intercept and 9 orthonormal polynomials), 10 random effects covariance parameters and $\sigma_e^2$ (assuming $\Sigma_{di}(\tau_d) = \text{diag}(\tau_d)$ and $\Sigma_{ei}(\tau_e) = \sigma_e^2 I_{p_i}$) for a total of 11 covariance parameters. We also compare the fit using $\Sigma_{di}(\tau_d) = \text{diag}(\tau_d)$ to that using an unstructured covariance for $\Sigma_{di}(\tau_d)$. However, with conventional mixed effects software like SAS Proc Mixed, it would be impossible to use an unstructured $\Sigma_{di}(\tau_d)$ with natural polynomials for the subject-specific models for the reasons discussed in section 3.1.

*3.3 The Case of Incomplete, Unbalanced Data*

In the DASH Study, about 81% (288/357) of the subjects had complete data at baseline. The additional 19% of subjects had very little missing data. We assumed that the data were missing competely at random for these analyses, i.e., an "intent-to-treat" treatment of missing data by including every observation of the dependent measure (no observations are discarded and no data are imputed).

One of the main advantages of the linear mixed model for longitudinal data is that it uses all available data, rather than excluding cases with missing observations, and accommodates unbalanced designs. This feature of the linear mixed model is of particular relevance to ABPM data since it is common for a significant proportion of participants to be missing at least one observation.[27]

In 24-hour ABPM studies, typically there are criteria for controlling the quality of the monitoring data. For example, in the DASH study, if fewer than 14 acceptable readings were obtained, the subject was asked to repeat the monitoring. Among participants with acceptable ABPM measurements, more than 90% of the possible waking and sleeping readings were obtained.[28] Sometimes more rigid criteria are used, e.g., for Hermida et al.[29], BP series were not considered valid for analysis if 30% of the measurements were lacking, if data were missing for >2-hour spans, if data were collected from subjects while experiencing an irregular rest–activity schedule, or if the nighttime sleep span was <6 hours or >12 hours during monitoring.

## 4. DASH Baseline Data Analysis Results - Estimation, Inference and Goodness-of-fit

Table 1 provides estimates, standard errors (SE), and p-values for the 9th degree orthonormal polynomial fit to the DASH baseline 24-hour ABPM data. Table 2 provides model comparisons using ornothormal polynomial models of degree ranging 4-9. All computations were done using SAS v9.2 and the *ORPOL* function in the SAS *IML* procedure, which handles equal and unequal time spacings when generating orthonormal polynomials. Restricted maximum likelihood estimation (REML) was used for estimation and the Kenward-Roger F and adjusted denominator degrees of freedom were used for all fixed effect inference. However, maximum likelihood estimation results were almost identical (results not shown). From Table 1, we can see that the sizes of the absolute values of the orthonormal polynomial regression coefficients indicate that orthonormal polynomials 1-6 have the largest effect on BP, changing from being greater than 4 to less than 2.

Table 2 provides model selection criteria results for the Akaike Information Criterion[30] (AIC) and Bayesian Information Criterion[31] for orthonormal polynomial models using 4-9th degree polynomials in both the fixed and random effects. AIC and BIC were computed for both $\Sigma_{di}(\tau_d) = \text{diag}(\tau_d)$ and $\Sigma_{di}(\tau_d) = $ Unstructured (UN). Morrell et al.[32] commented that the best way to select among linear mixed-effects models based on various information criteria is still not clearly determined. The conclusion from both the AIC and BIC is that the 9th degree orthonormal polynomial model using $\Sigma_{di}(\tau_d) = $ UN is the best model for DBP and SBP since it yields the smallest AIC and BIC values when compared to all other orthonormal polynomial models. Mathematically, the assumption of $\Sigma_{di}(\tau_d) = \text{diag}(\tau_d)$ should have yielded the best covariance fit, but the lack of fit in the fixed effects is carried over into the variance (the estimated $\Sigma_{di}(\tau_d)$ was not diagonal). There were no model convergence problems using $\Sigma_{di}(\tau_d) = $ UN.

To check the model distribution assumptions for the random effects and residual error, we used an important result from Gurka et al.[33] which states that if $\epsilon_i = Z_i d_i + e_i$ is multivariate normal, then both $d_i$ and $e_i$ are multivariate normal given the model assumption that $d_i$ and $e_i$ are independent. Using this result, we used standard residual analysis for the estimated "stacked" $\epsilon_i$ and concluded that the errors were approximately normally distributed for each of the orthonormal polynomial models considered.

The linear mixed model easily accommodates additional explanatory variables. For example, we can do a simple check for baseline homogeneity of average 24-hour blood pressure for the 3 diet groups by including intercept effects for diet (two indicator variables with control diet as the reference group) and interaction effects in time for diet. Let $\delta_i = 1$ if subject $i$ is assigned the diet rich in fruits and vegetables and $\delta_i = 0$ otherwise. Similarly, let $\lambda_i = 1$ if subject $i$ is assigned the DASH diet and $\lambda_i = 0$ otherwise. We can then add diet fixed effects (both main effects and diet x time interaction effects) in the linear mixed model as follows

$$\boldsymbol{y}_i = (\, \boldsymbol{X}_i^* \quad \delta_i \boldsymbol{X}_i^* \quad \lambda_i \boldsymbol{X}_i^* \,) \begin{pmatrix} \boldsymbol{\beta} \\ \boldsymbol{\alpha} \\ \boldsymbol{\gamma} \end{pmatrix} + \boldsymbol{Z}_i^* \boldsymbol{d}_i + \boldsymbol{e}_i \,. \tag{5}$$

Here $\boldsymbol{\beta}$, $\boldsymbol{\alpha}$, and $\boldsymbol{\gamma}$ are $q \times 1$ vectors of fixed effect population parameters where $\boldsymbol{\beta}$ represents the effect of the control diet (intercept and time) on blood pressure, $\boldsymbol{\alpha}$ represents the differential effect of the diet rich in fruits and vegetables relative to the control diet, and $\boldsymbol{\gamma}$ represents the differential effect of the DASH diet relative to the control diet. Note that the structure of random effects and within-subject errors remain the same as in equation (4) as does the corresponding covariance structure. We can rewrite the model as a standard linear mixed model

$$\boldsymbol{y}_i = \boldsymbol{X}_i \boldsymbol{\beta} + \boldsymbol{Z}_i \boldsymbol{d}_i + \boldsymbol{e}_i \,. \tag{6}$$

Here $\boldsymbol{X}_i$ is a $p_i \times 3q$ fixed effects design matrix for person $i$ and $\boldsymbol{\beta}$ is the corresponding $3q \times 1$ vector of fixed effect parameters. Tests of interaction and intercept effects (not shown here) were not signifcant and hence we concluded that there was no evidence of a difference in 24-hour BP profiles at baseline. Using the same model for the intervention period, there was no evidence of interaction effects but there was evidence of diet main effects (again, results not shown here). An example of including several explanatory variables in the analysis of the DASH data can be found in Simpson and Edwards[26], which features a general linear model for longitudinal data with the circular LEAR correlation structure and uses diet groups, race, age, and orthonormal polynomials as fixed effects.

In order to incorporate both baseline and intervention into the model, we use a shared parameter, multivariate linear mixed model[34-35] given by

$$\begin{pmatrix} \boldsymbol{y}_{0i} \\ \boldsymbol{y}_{1i} \end{pmatrix} = \begin{pmatrix} \boldsymbol{X}_{0i} \\ \boldsymbol{X}_{1i} \end{pmatrix} \boldsymbol{\beta} + \begin{pmatrix} \boldsymbol{Z}_{0i} \\ \boldsymbol{Z}_{1i} \end{pmatrix} \boldsymbol{d}_i + \begin{pmatrix} \boldsymbol{e}_{0i} \\ \boldsymbol{e}_{1i} \end{pmatrix}, \qquad (7)$$

where $\boldsymbol{y}_{0i}$ is the $24 \times 1$ vector of observations for baseline BP, $\boldsymbol{y}_{1i}$ is the $24 \times 1$ vector of observations for intervention BP, $\boldsymbol{X}_{0i}$ and $\boldsymbol{X}_{1i}$ the $24 \times q$ fixed effects design matrices for baseline and intervention periods, and $\boldsymbol{Z}_{0i}$ and $\boldsymbol{Z}_{1i}$ the $24 \times m$ random effects design matrices for baseline and intervention periods. The shared parameter model is appropriate since the response is the same for baseline and intervention. The scalar representation of the model is given by

$$y_{hij} = \beta_0 + \beta_1 \pi_{hi} + \beta_2 \delta_i + \beta_3 \lambda_i + \beta_4 \pi_{hi} \delta_i + \beta_5 \pi_{hi} \lambda_i + \sum_{k=6}^{14} \beta_k x^*_{kij} + \sum_{l=1}^{10} d_{il} z^*_{lij} + e_{hij}$$

where $h = 0,1$ represents baseline and intervention BP for subject $i$, $\pi_{0i} = 0$ if baseline period and $\pi_{1i} = 1$ if intervention period for subject $i$, and $x^*_{kij}$ and $z^*_{lij}$ are the fixed and random orthonormal time effects. The model assumes that within-subject variation is the same at baseline and intervention and the between-subject effect of intervention is modeled using the period effect, $\pi_{hi}$, for the intercept and diet main effects ($\beta_1 \pi_{hi}$, $\beta_4 \pi_{hi} \delta_i$, $\beta_5 \pi_{hi} \lambda_i$). Table 3 provides the results for the change in SBP and DBP for each diet main effect. Both the DASH and fruit/vegetable diets lowered SBP and DBP significantly compared with the control diet ($p < 0.0001$ for SBP and DBP on both diets: control diet, $-0.3/-0.02$ mm Hg; fruit/vegetable diet, $-3.2/-2.1$ mm Hg; DASH diet, $-4.6/-2.8$ mm Hg). The model results are very similar to the results Moore et al. (1999 - 28) found using mean ABP where both the fruit/vegetable and DASH diets lowered 24-hour ABP significantly compared with the control diet (p < 0.0001 for SBP and DBP on both diets: control diet, $-0.2/+0.1$ mm Hg; fruit/vegetable diet, $-3.2/-1.9$ mm Hg; DASH diet, $-4.6/-2.6$ mm Hg).

## 5. Model-Based Graphical Displays for Evaluating 24-hour ABPM Data and AUC

*5.1 Graphical Displays*

As discussed in section 1, O'Brien[5] posed the important question regarding how the data should be presented. O'Brien[5] provided depictions of 24-hour ABPM data based on observed values that contained a plot of normal bands across the 24 hours for which he overlayed selected individual subject trajectories to demonstrate examples of normal blood pressure and various instances of deviations from normal blood pressure (above or below normal range). O'Brien[5] noted that as with conventional measurement, normal ranges for ABPM have been the subject of much debate over the years. The focus of the outcomes-based figures is to provide practicing physicians with a graphical approach to evaluating an individual's 24-hour BP. However, for clinical and observational studies that require group comparisons and exploring additional explanatory factors that may impact BP, the outcomes-based graphical display is not as useful. A model-based approach helps overcome some of the limitations of the outcomes-based approach.

The linear mixed model with orthonormal polynomials provides the statistical framework for producing a model-based graphical display which will be extremely useful for clinical and observational studies of 24-hour ABPM. Subject-specific trajectories can be of great importance in ABPM studies, especially in clinical practice and research. In practice, it is the subject-specific values that are used to provide diagnoses and to design treatment regimens. However, little emphasis has been placed on this dimension of the problem in modeling 24-hour ABPM data. The smoothed predicted trajectories will provide researchers with the ability to construct additional measures of 24-hour ABPM and correlate the measures with important cardiovascular disease (CVD) outcomes such as hypertension and electrocardiogram measures such as left ventricle mass index (LVMI).

Figures 1 and 2 for SBP and DBP provide model-based graphical displays of 24-hour ABPM based on O'Brien[2] outcomes-based concepts. Figures 1 and 2 show one subject's predicted curve computed from the linear mixed model for SBP and DBP (solid blue lines), the subject's observed data (connected by dashed blue lines), the 90% prediction interval for DASH study normals at baseline (shaded gray region), and the commonly used static measures of upper limit of normal for daytime and nighttime (red horizontal lines) SBP (135/120) and DBP (85/75). The 90% prediction interval for the 9th degree orthonormal polynomial model provides a model-based definition of normal range. The 90% prediction interval closely matches the outcomes-based $\pm 2$ standard deviations about the mean for each time point. We constructed the 90% prediction interval by selecting the subset of all subjects that had normal SBP (DBP) at baseline, fitting a linear mixed model with orthonormal polynomials, and then computing a 90% prediction interval using a SAS macro developed by By[36]. The prediction intervals are used to provide an interval estimate for a single observation as opposed to confidence intervals which provide an (narrower) interval estimate for an average of observations.

For SBP, the smoothed subject-specific trajectory suggests that the subject is above the upper limit of normal for the majority of the day and night. However, there are observed values for which the smoothed subject-specific trajectory suggests there may be masked hypertension - 10:00-11:00 a.m. and 1:00-4:00 p.m., hours 10-11 and 13-16 in Figure 1, respectively. For DBP, both the smoothed subject-specific trajectory and the observed data indicate that this person's DBP is above the upper limit of normal for both daytime and nighttime. However, in practice, actual observed BP values will necessarily take precedence over predicted values.

The graphical displays provided in Figures 1 and 2 also visually reveal that the discontinuity at the change point from daytime to nighttime can present a problem for assessing observed values and for computing AUC values for the smoothed subject-specific trajectories using the static cutoffs. It is clear that the static cutoffs do not take into account the natural variability and downward trend of the 24-hour ABPM data near the discontinuity.

*5.2 Area Under the Curve (AUC)*

The DASH diet is now well known to reduce blood pressure. Thus, model-based subject-specific AUC should correlate with the perfromance of the DASH diet. Table 3 provides a descriptive comparison of the average model-based subject-specific AUC for the control, DASH, and fruits/vegetable diets. Results are for the 288 subjects who had complete 24-ABPM data at baseline and the 287 subjects who had complete data during the intervention. The DASH diet group clearly has the largest drop from baseline of average AUC for both SBP and DBP, followed by the fruits/vegetables group and then the control group having the poorest performance. For the DASH diet, the mean change in the model based AUC divided by 24 hours for SBP is 5.2 mmHg and DBP is 3.0 mmHg. The model-based AUC result is consistent with the DASH study finding that the DASH diet lowered blood pressures by an average of 5.5 and 3.0 mmHg for systolic and diastolic, compared with the control diet.[37] Sacks et al.[37] DASH analyses defined baseline blood pressure as the average of 7 days of measurements, 3 during screening and 4 during run-in. The last five sets of measurements during the intervention were used to compute end of study blood pressure, and the effect of the diets. For mean change in 24-hour ABPM measurements, Sacks et al.[37] reported 4.5 mmHg for SBP and 2.7 mmHg for DBP.

The model-based subject-specific scaled AUC can be shown to be a linear combination of the fixed and random effect regression coefficients. In order to demonstrate, suppose that we were using natural polynomials, then equation 4 assumes $\boldsymbol{X}_i^* = \boldsymbol{Z}^*$ and that BP is a polynomial function in time, $x$, and hence the subject-specific AUC is given by

$$\text{AUC}_i = \int_0^{24} (\widehat{\beta}_{i0} + \sum_{j=1}^{9} \widehat{\beta}_{ij} x^j) dx = 24\widehat{\beta}_{i0} + \sum_{j=1}^{9} \widehat{\beta}_{ij} \frac{24^{j+1}}{j+1}$$

and the corresponding AUC scaled by 24 hours is

$$\text{AUC}_i/24 = \widehat{\beta}_{i0} + \sum_{j=1}^{9} \frac{24^j}{j+1} \widehat{\beta}_{ij},$$

where $\widehat{\beta}_{i0} = \widehat{\beta}_0 + \widehat{d}_{i0}$ is the estimated BP intercept for the $i$-th subject and $\widehat{\beta}_{ij} = \widehat{\beta}_j + \widehat{d}_{ij}, j > 0$, is the estimated $j$-th regression coefficient for the $i$-th subject. So the model-based subject-specific scaled AUC is a linear combination of the adjusted mean BP (subject-specific intercept) and a weighted linear combination of subject-specific regression coefficients. It is clear that the scaled AUC is different than a simple mean of 24-hour ABPM data. However, further research is required to assess whether the model-based subject-specific AUC is a better summary measure of 24-hour ABPM than the simple arithmetic mean that is typically used.

These results have very important implications for the use of model-based subject-specific AUC computed from 24-ABPM data in CVD studies. They provide some evidence that model-based AUC computed from 24-ABPM data may potentially be used as a correlate of CVD outcomes. For example, Nobre and Mion[38] calculated areas under systolic and diastolic blood pressure trajectories using the observed data (outcomes-based) and compared with systolic and diastolic blood pressure load and 24-hour systolic and diastolic blood pressure in order to determine which provided the best correlation with left ventricular mass index (LVMI). Nobre and Mion[38] found that the correlations of the parameters obtained by ABPM with LVMI were high and statistically significant except for blood pressure load between 90 and 100% and for 24-hour SBP of 135 mmHg or less, and SBP load higher than 50%.

## 6. Discussion

The development of alternative, robust and flexible statistical methods for analyzing 24-hour ABPM data will be important in refining its role in clinical research, clinical practice, and observational studies. Therefore, we were motivated to develop an alternative statistical model for analyzing 24-hour ABPM data that will overcome some of the limitations of conventional analyses and potentially lead to improved 24-hour ABPM correlates of CVD outcomes and electrocariogram measures such as LVMI. Though the DASH study did not measure electrocardiogram parameters, the Jackson Heart Study[39] is an example of a study that has both baseline 24-hour ABPM and LVMI neasurements.

We have shown how the linear mixed model with orthonormal polynomials across time in both the fixed and random effects provides a powerful approach to the analysis of 24-hour ABPM data, particularly when focus is on modeling the subject-specific trajectories. In addition to replicating DASH Study results, the linear mixed model with orthonormal polynomials provides researchers with the ability to explore the simultaneous effects of multiple predictors and their interactions. Because data visualization can be a very useful tool for assessing 24-hour ABPM, we presented model-based graphical presentations of 24-hour ABPM based on the linear mixed model results that can aid in the visual representation and interpretation of 24-hour ABPM data. However, further research will be required to fully understand the robustness of the orthonormal polynomial model to outliers, departures from distributional assumptions, and the effect of more severe missingness, irregularly timed data, and truncated data. In addition, different assumptions and parameterizations for the multivariate linear mixed model should be further explored in order to gain further insight into the statistical properties of the differences between DASH Study 24-hour ABP at baseline and intervention.

The linear mixed model is relatively easy to implement (given the complexity of the technique) using available software, allows for straight-forward testing of multiple hypotheses, and the results can be communicated effectively to research clinicians using both graphical and tabular displays. Using orthonormal polynomials provides the ability to model the nonlinear trajectories of each subject with the same complexity as the mean model (fixed effects).

**Acknowledgements**

Lloyd Edwards was supported by the National Center for Research Resources and the National Center for Advancing Translational Sciences, National Institutes of Health, through Grant Award Number UL1TR000083. Sean Simpson was supported by NIBIB K25 EB012236-01A1.

REFERENCES

1. Appel LJ and Stason WB. Ambulatory blood pressure monitoring and blood pressure self-measurement in the diagnosis and management of hypertension. *Ann Intern Med* 1993; 118: 889-892.

2. O'Brien E. Ambulatory blood pressure measurement: the case for implementation in primary care. *Hypertension* 2008; 51: 1435-1441.

3. O'Brien E. Ambulatory blood pressure monitoring: 24 hour blood pressure control as a therapeutic goal for improving cardiovascular prognosis. *Medicographia* 2010; 32: 241-249 .

4. Owens P, Atkins N, O'Brien E. Diagnosis of white coat hypertension by ambulatory blood pressure monitoring. *Hypertension* 1999; 34: 267–72.

5. O'Brien E. Ambulatory blood pressure monitoring in the management of hypertension. *Heart* 2003; 89: 571-576.

6. Lambert PC, Abrams KR, Jones DR, et al. Analysis of ambulatory blood pressure monitor data using a hierarchical model incorporating restricted cubic splines and heterogeneous within-subject variances. *Stat Med* 2001; 20: 3789-3805.

7. Weber MA, Cheung DG, Grarttinger WF. et al. Characterization of antihypertensive therapy by whole-day blood pressure monitoring. *JAMA* 1988; 259: 3281-3285.

8. Ferguson JH and Shaar CJ. The effective diagnosis and treatment of hypertension by the primary care physician: impact of ambulatory blood pressure monitoring. *J Am Board Fam Pract* 1992; 5: 457-65.

9. Appel LJ, Moore TJ, Obarzanek E, et al. A clinical trial of the effects of dietary patterns on blood pressure. DASH collaborative research group. *N Engl J Med* 1997; 336: 1117-1124.

10. Halberg J, Halberg F, Leach CN. Variability of human blood pressure with reference mostly to the non-chronologic literature. *Chronobiologia* 1984; 11: 205-16.

11. Gaffney M, Taylor C, Cusenza E. Harmonic regression analysis of the effect of drug treatment on the diurnal rhythm of blood pressure and angina. *Stat Med* 1993; 12: 129-142.

12. Gaffney M, Taylor C, Cusenza E. Harmonic regression analysis of the effect of drug treatment on the diurnal rhythm of blood pressure and angina. *Stat Med* 1993; 12:129-142.

13. Streitberg B, Mayer-Sabellek W, Baumgart P. Statistical analysis of circadian blood pressure recordings in controlled clinical trials. *J Hypertens Suppl* 1989; 7: S11-S17.

14. Streitberg, B., Mayer-Sabellek, W. Smoothing twenty-four-hour ambulatory blood pressure profiles: a comparison of alternative methods. *J Hypertens Suppl* (1990; 8: S21-S37.

15. Dickson D and Hasford J. 24-hour blood pressure measurement in antihypertensive drug trials: Data requirements and methods of Analysis. *Stat Med* 1992; 11: 2147-2158.

16. Jaccard J, Wan CK. Statistical analysis of temporal data with many observations: issues for behavioral medicine data. *Ann Behav Med* 1993; 15: 41-50.

17. Schwartz JE, Warren K, Pickering TG. Mood, location and physical position as predictors of ambulatory blood pressure and heart rate: Application of a multi-level random effects model. *Ann Behav Med* 1994; 16: 210-220.

18. Selwyn MR and Difranco DM. The application of large Gaussian mixed models to the analysis of 24 hour ambulatory blood pressure monitoring data in clinical trials. *Stat Med* 1993; 12: 1665-1682.

19. Harville DA. Extension of the Gauss-Markov theorem to include the estimation of random effects. *Ann Stat* 1976; 4: 384–395.

20. Laird NM and Ware JH. Random-effects models for longitudinal data. *Biometrics* 1982; 38: 963-974.

21. Edward LJ. Modern statistical techniques for the analysis of longitudinal data in biomedical research. Pediatr Pulmonol 2000; 30: 330-334

22. Gurka MJ and Edwards LJ. Mixed Models. *Handbook of Statistics, Volume 27: Epidemiology and Medical Statistics*. Elsevier, North-Holland: Amsterdam 2007, pp. 253–280.

23. Cheng J, Edwards LJ, Maldonado-Molina MM, et al. Real longitudinal data analysis for real people: Building a good enough mixed model. *Stat Med* 2010; 29: 504-520.

24. Jeffreys H and Jeffreys BS. Weierstrass's theorem on approximation by polynomials and extension of Weierstrass's approximation theory. *Methods of Mathematical Physics*, 3rd ed. Cambridge, England: Cambridge University Press, 1988, pp. 446-448.

25. Edwards LJ, Stewart PJ, MacDougall JE, et al. A method for fitting regression splines with varying polynomial order in the linear mixed model. *Stat in Med* 2006; 25: 513-527.

26. Simpson SL, Edwards LJ. A circular LEAR correlation structure for cyclical longitudinal data. *Stat Methods Med Res* 2011; D0I:10.1177/0962280210395741).

27. Ibrahim JG and Molenberghs G. Missing data methods in longitudinal studies: a review. *Test (Madr)* 2009; 8: 1-43.

28. Moore TJ, Vollmer WM, Appel LJ, et al. Effect of dietary patterns on ambulatory blood pressure: results from the Dietary Approaches to Stop Hypertension (DASH) trial. DASH Collaborative Research Group. *Hypertension* 1999; 34: 472-477.

29. Hermida RC, Ayala DE, Calvo C, et al. Effects of time of day of treatment on ambulatory blood pressure pattern of patients with resistant hypertension. *Hypertension* 2005; 46: 1053–1059.

30. Akaike H. A new look at the statistical model identification. *IEEE Trans Automatic Control* 1974; AC-19: 716-723.

31. Schwarz SR. Estimating the dimension of a model. *Ann Stat* 1978; 6: 461-464.

32. Morrell CH, Brant LJ, Ferrucci L. Model choice can obscure results in longitudinal studies. *J Gerontol A Biol Sci Med Sci* 2009; 64: 215–222.

33. Gurka MJ, Edwards LJ, Muller KE, et al. Extending the Box–Cox transformation to the linear mixed model. *J R Stat Soc Ser A Stat Soc* 2006; 169: 273–288.

34. Verbeke G, Fieuws S, Molenberghs G. et al. The analysis of multivariate longitudinal data: A review. *Stat Methods Med Res* 2012; DOI 10.1177/0962280212445834.

35. Bandyopadhyay S, Ganguli B, Chatterjee A. A review of multivariate longitudinal data analysis. *Stat Methods Med Res* 2011; 20: 299-330.

36. By K. Macro for computing prediction interval. http://www.unc.edu/~kby/sas/predVar.sas (2005).

37. Sacks FM, Appel LJ, Moore TJ, et al. The Dietary Approach to Prevent Hypertension: A review of the Dietary Approaches to Stop Hypertension (DASH) Study. *Clin Cardiol* 1999; 22: (Suppl. III), III-6-III-10.

38. Nobre F and Mion D. Is the area under blood pressure curve the best parameter to evaluate 24-h ambulatory blood pressure monitoring data? *Blood Press Monit* 2005; 10: 263–270.

39. Carpenter MA, Crow R, Steffes M, et al. Laboratory, Reading Center, and Coordinating Center Data management Methods in the Jackson Heart Study. *Am J Med Sci* 2004; 328: 131-144.

**Table 1. Fixed Effect Estimates, Standard Errors (SE), and P-values for 9th Degree Orthornormal Polynomial Fit to Baseline DASH Study Data, N = 357**

| Outcome | Parameter | Ortho Poly Degree | Estimate | SE | P-value |
|---|---|---|---|---|---|
| DBP | $\beta_0$ | Intercept | 83.71 | 0.394 | < 0.0001 |
| | $\beta_1$ | 1 | − 22.98 | 0.795 | < 0.0001 |
| | $\beta_2$ | 2 | − 13.41 | 0.748 | < 0.0001 |
| | $\beta_3$ | 3 | 9.91 | 0.605 | < 0.0001 |
| | $\beta_4$ | 4 | 9.61 | 0.584 | < 0.0001 |
| | $\beta_5$ | 5 | 4.35 | 0.518 | < 0.0001 |
| | $\beta_6$ | 6 | − 4.62 | 0.436 | < 0.0001 |
| | $\beta_7$ | 7 | − 0.51 | 0.471 | 0.2768 |
| | $\beta_8$ | 8 | 1.96 | 0.420 | < 0.0001 |
| | $\beta_9$ | 9 | 0.58 | 0.428 | 0.1782 |
| SBP | $\beta_0$ | Intercept | 131.64 | 0.575 | < 0.0001 |
| | $\beta_1$ | 1 | − 25.03 | 0.972 | < 0.0001 |
| | $\beta_2$ | 2 | − 19.71 | 0.909 | < 0.0001 |
| | $\beta_3$ | 3 | 10.18 | 0.732 | < 0.0001 |
| | $\beta_4$ | 4 | 10.74 | 0.708 | < 0.0001 |
| | $\beta_5$ | 5 | 5.19 | 0.614 | < 0.0001 |
| | $\beta_6$ | 6 | − 4.38 | 0.528 | < 0.0001 |
| | $\beta_7$ | 7 | − 0.74 | 0.542 | 0.1722 |
| | $\beta_8$ | 8 | 1.13 | 0.466 | 0.0153 |
| | $\beta_9$ | 9 | 1.43 | 0.470 | 0.0025 |

**Table 2. Model Comparisons: AIC and BIC, FE=Fixed Effects, RE=Random Effects, Baseline DASH Study Data, N = 357**

| | | $\Sigma_{di}(\tau_d) = \text{diag}(\tau_d)$ | | $\Sigma_{di}(\tau_d) = \text{UN}$ | |
|---|---|---|---|---|---|
| Outcome | Model | AIC | BIC | AIC | BIC |
| DBP | Ortho Poly (FE_degree = 9, RE_degree = 9) | 58,989 | 59,032 | 58,710 | 58,927 |
| | Ortho Poly (FE_degree = 8, RE_degree = 8) | 59,031 | 59,070 | 58,784 | 58,963 |
| | Ortho Poly (FE_degree = 7, RE_degree = 7) | 59,089 | 59,124 | 58,873 | 59,016 |
| | Ortho Poly (FE_degree = 6, RE_degree = 6) | 59,158 | 59,189 | 58,962 | 59,074 |
| | Ortho Poly (FE_degree = 5, RE_degree = 5) | 59,344 | 59,371 | 59,180 | 59,266 |
| | Ortho Poly (FE_degree = 4, RE_degree = 4) | 59,570 | 59,593 | 59,502 | 59,564 |
| SBP | Ortho Poly (FE_degree = 9, RE_degree = 9) | 61,598 | 61,640 | 61,336 | 61,553 |
| | Ortho Poly (FE_degree = 8, RE_degree = 8) | 61,637 | 61,676 | 61,410 | 61,588 |
| | Ortho Poly (FE_degree = 7, RE_degree = 7) | 61,665 | 61,699 | 61,452 | 61,596 |
| | Ortho Poly (FE_degree = 6, RE_degree = 6) | 61,741 | 61,772 | 61,543 | 61,655 |
| | Ortho Poly (FE_degree = 5, RE_degree = 5) | 61,901 | 61,928 | 61,748 | 61,833 |
| | Ortho Poly (FE_degree = 4, RE_degree = 4) | 62,164 | 62,188 | 62,091 | 62,153 |

Table 3. Baseline and Intervention Diet Effects (p-values) for 9th Degree Orthornormal Polynomial Model Fit to DASH Data, N = 357

| | Change from Baseline to Intervention | | |
|---|---|---|---|
| Outcome | Control | Fruit/Veg | DASH |
| SBP | − 0.29 (0.1427) | − 3.23 ( < 0.0001) | − 4.55 ( < 0.0001) |
| DBP | − 0.02 (0.8906) | − 2.12 ( < 0.0001) | − 2.84 ( < 0.0001) |

Table 4. AUC Descriptive Statistics for SBP and DBP: Control, DASH, and Fruit/Veg, Baseline (N = 288) and Intervention (N = 287) DASH Study, Complete Data

|  |  | Baseline |  | Intervention |  | Mean Change |  |
| --- | --- | --- | --- | --- | --- | --- | --- |
| Diet | Outcome | N | Mean* (Std Err) | N | Mean* (Std Err) | AUC | AUC/24 |
| Control | SBP | 101 | 3026 (23.7) | 97 | 3013 (25.9) | 13 | 0.5 |
|  | DBP |  | 1923 (16.2) |  | 1928 (19.3) | −5 | 0.2 |
|  |  |  |  |  |  |  |  |
| DASH | SBP | 89 | 3045 (25.4) | 97 | 2920 (20.1) | 125 | 5.2 |
|  | DBP |  | 1923 (15.8) |  | 1852 (12.7) | 71 | 3.0 |
|  |  |  |  |  |  |  |  |
| Fruit/Veg | SBP | 98 | 3028 (21.4) | 93 | 2978 (24.6) | 50 | 2.1 |
|  | DBP |  | 1927 (17.1) |  | 1888 (17.8) | 39 | 1.6 |
|  |  |  |  |  |  |  |  |
| Total | SBP | 288 | 3032 (13.5) | 287 | 2970 (13.8) | 62 | 2.6 |
|  | DBP |  | 1924 (9.4) |  | 1889 (9.8) | 35 | 1.5 |

*Mean of subject-specific AUC computed from LMM

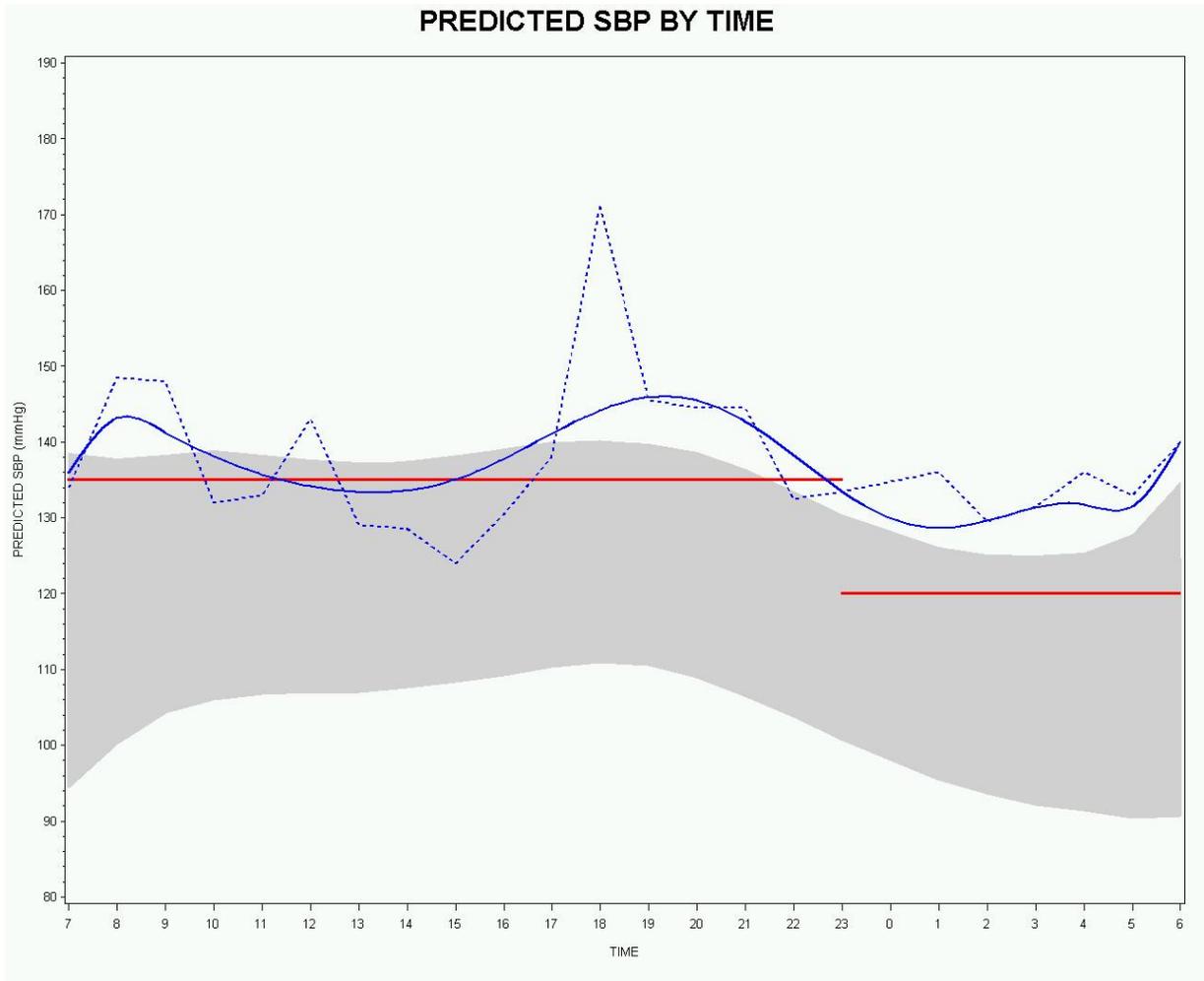

Figure 1. Predicted (blue solid curve) and observed (blue dotted curve) SBP by time for 1 subject with the 90% prediction interval (shaded region) based on normal subjects and commonly used upper normal limit static measures for daytime and nighttime (red horizontal lines).

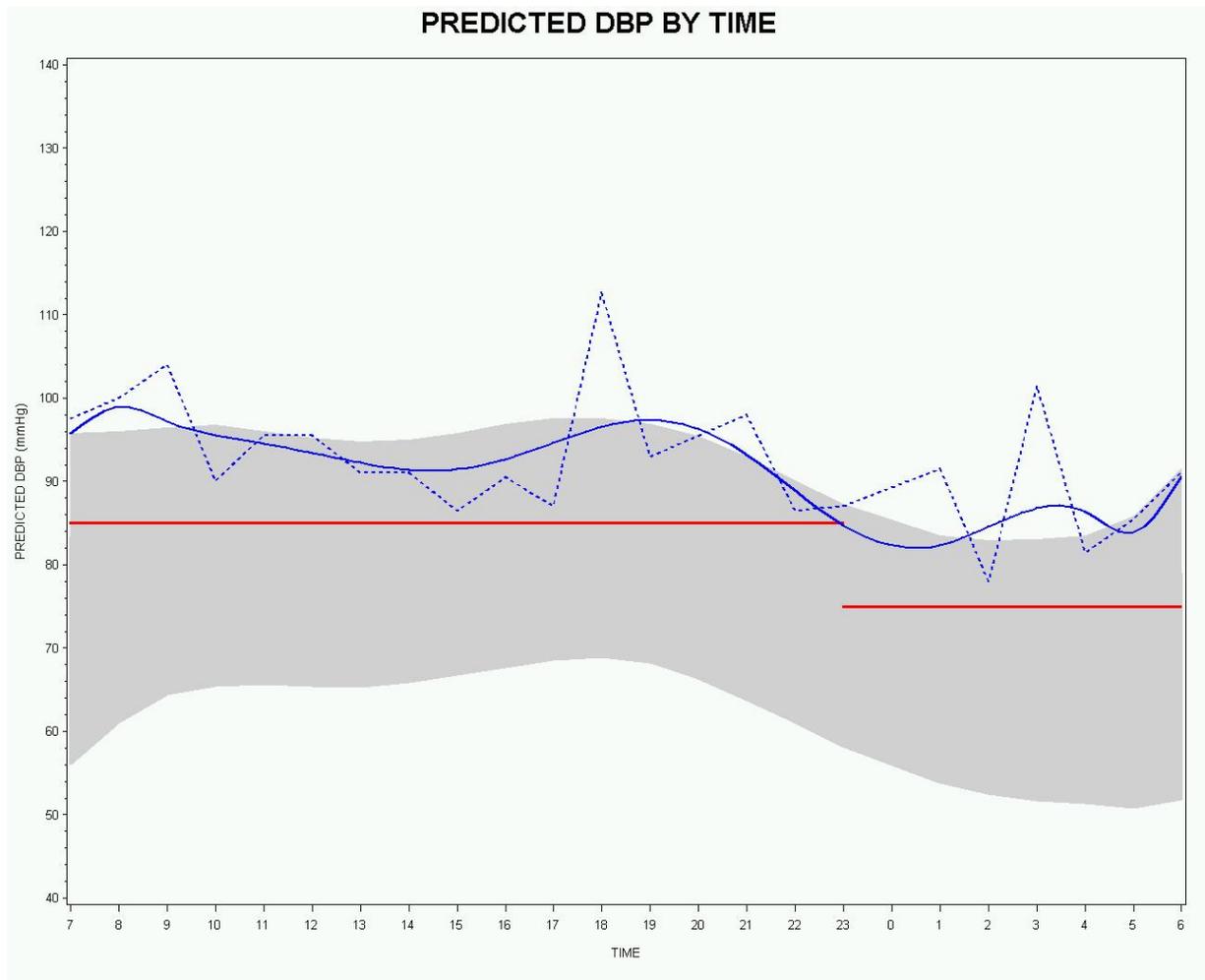

Figure 2. Predicted (blue solid curve) and observed (blue dotted curve) DBP by time for 1 subject with the 90% prediction interval (shaded region) based on normal subjects and commonly used upper normal limit static measures for daytime and nighttime (red horizontal lines).